\documentclass[twocolumn,prc,showpacs,superscriptaddress]{revtex4-1}

\usepackage{wasysym}
\usepackage{ulem}
\usepackage{graphicx}
\usepackage{dcolumn}
\newcolumntype{(}{D{(}{(}{-1}}

\begin{document}


\title{Constraining nova observables: direct measurements of resonance strengths in $^{33}$S($p,\gamma$)$^{34}$Cl}


\author{J.~Fallis}\email{jfallis@trumf.ca}
\affiliation{TRIUMF, Vancouver, British Columbia  V6T 2A3, Canada}
\author{A.~Parikh}\affiliation{Departament de F\'{i}sica i Enginyeria Nuclear, EUETIB, Universitat Polit\`{e}cnica de Catalunya, c/ Comte {d'Urgell} 187,
E-08036 Barcelona, Spain}\affiliation{Institut {d'Estudis} Espacials de Catalunya, c/Gran Capita 2-4, Ed. Nexus-201, E-08034 Barcelona, Spain}\affiliation{Physik Department E12, Technische Universit\"{a}t M\"{u}nchen, D-85748 Garching, Germany}
\author{P.$\,$F.~Bertone}\affiliation{Physics Division, Argonne National Laboratory, Argonne, IL 60439, USA}
\author{S.~Bishop}\affiliation{Physik Department E12, Technische Universit\"{a}t M\"{u}nchen, D-85748 Garching, Germany}
\author{L.~Buchmann}\affiliation{TRIUMF, Vancouver, British Columbia  V6T 2A3, Canada}
\author{A.$\,$A.~Chen }\affiliation{McMaster University, Hamilton, ON, Canada}
\author{G.~Christian}\affiliation{TRIUMF, Vancouver, British Columbia  V6T 2A3, Canada}
\author{J.$\,$A.~Clark}\affiliation{Physics Division, Argonne National Laboratory, Argonne, IL 60439, USA}
\author{ J.$\,$M.~{D'Auria}}\affiliation{Department of Physics, Simon Fraser University, Burnaby, BC V5A 1S6, Canada}
\author{B.~Davids}\affiliation{TRIUMF, Vancouver, British Columbia  V6T 2A3, Canada}\affiliation{Department of Physics, Simon Fraser University, Burnaby, BC V5A 1S6, Canada}
\author{C.$\,$M.~Deibel}\affiliation{Physics Division, Argonne National Laboratory, Argonne, IL 60439, USA}\affiliation{Joint Institute for Nuclear Astrophysics, Michigan State University, East Lansing, MI 48824, USA}\affiliation{Department of Physics and Astronomy, Louisiana State University, Baton Rouge, LA 70803, USA}
\author{B.$\,$R.~Fulton}\affiliation{ Department of Physics, University of York, York, YO10 5DD, UK}
\author{U.~Greife}\affiliation{Colorado School of Mines, Golden, CO 80401, USA}
\author{B.~Guo}\affiliation{China Institute of Atomic Energy, Beijing 102413, People's Republic of China}
\author{U.~Hager}\affiliation{TRIUMF, Vancouver, British Columbia  V6T 2A3, Canada}\affiliation{Colorado School of Mines, Golden, CO 80401, USA}
\author{C.~Herlitzius}\affiliation{Physik Department E12, Technische Universit\"{a}t M\"{u}nchen, D-85748 Garching, Germany}
\author{D.$\,$A.~Hutcheon}\affiliation{TRIUMF, Vancouver, British Columbia  V6T 2A3, Canada}
\author{ J.~Jos\'{e}}\affiliation{Departament de F\'{i}sica i Enginyeria Nuclear, EUETIB, Universitat Polit\`{e}cnica de Catalunya, c/ Comte {d'Urgell} 187,
E-08036 Barcelona, Spain}\affiliation{Institut {d'Estudis} Espacials de Catalunya, c/Gran Capita 2-4, Ed. Nexus-201, E-08034 Barcelona, Spain}
\author{A.$\,$M.~Laird}\affiliation{ Department of Physics, University of York, York, YO10 5DD, UK}
\author{E.$\,$T.~Li}\affiliation{China Institute of Atomic Energy, Beijing 102413, People's Republic of China}
\author{Z.$\,$H.~Li}\affiliation{China Institute of Atomic Energy, Beijing 102413, People's Republic of China}
\author{G.~Lian}\affiliation{China Institute of Atomic Energy, Beijing 102413, People's Republic of China}
\author{W.$\,$P.~Liu}\affiliation{China Institute of Atomic Energy, Beijing 102413, People's Republic of China}
\author{L.~Martin}\affiliation{TRIUMF, Vancouver, British Columbia  V6T 2A3, Canada}
\author{K.~Nelson}\affiliation{TRIUMF, Vancouver, British Columbia  V6T 2A3, Canada}\affiliation{McMaster University, Hamilton, ON, Canada}
\author{D.~Ottewell}\affiliation{TRIUMF, Vancouver, British Columbia  V6T 2A3, Canada}
\author{P.$\,$D.~Parker}\affiliation{Wright Nuclear Structure Laboratory, Yale University, New Haven, Connecticut 06520, USA}
\author{S.~Reeve}\affiliation{TRIUMF, Vancouver, British Columbia  V6T 2A3, Canada}\affiliation{Department of Physics, Simon Fraser University, Burnaby, BC V5A 1S6, Canada}
\author{A.~Rojas}\affiliation{TRIUMF, Vancouver, British Columbia  V6T 2A3, Canada}
\author{C.~Ruiz}\affiliation{TRIUMF, Vancouver, British Columbia  V6T 2A3, Canada}

\author{K.~Setoodehnia}\affiliation{McMaster University, Hamilton, ON, Canada}
\author{S.~Sjue}\affiliation{TRIUMF, Vancouver, British Columbia  V6T 2A3, Canada}
\author{C.~Vockenhuber}\affiliation{ETH Zurich, Zurich, Switzerland}
\author{Y.$\,$B.~Wang}\affiliation{China Institute of Atomic Energy, Beijing 102413, People's Republic of China}
\author{C.~Wrede}\affiliation{Department of Physics, University of Washington, Seattle, WA 98195-1560, USA}\affiliation{Department of Physics and Astronomy and National Superconducting Cyclotron Laboratory, Michigan State University, East Lansing, Michigan 48824, USA}



\date{\today}

\begin{abstract}
The $^{33}$S($p,\gamma$)$^{34}$Cl reaction is important for constraining predictions of certain isotopic abundances in oxygen-neon novae.  Models currently predict as much as 150 times the solar abundance of $^{33}$S in oxygen-neon nova ejecta. This overproduction factor may, however, vary by orders of magnitude due to uncertainties in the $^{33}$S($p,\gamma$)$^{34}$Cl reaction rate at nova peak temperatures.  Depending on this rate, $^{33}$S could potentially be used as a diagnostic tool for classifying certain types of presolar grains. Better knowledge of the $^{33}$S($p,\gamma$)$^{34}$Cl rate would also aid in interpreting nova observations over the S-Ca mass region and contribute to the firm establishment of the maximum endpoint of nova nucleosynthesis. Additionally, the total S elemental abundance which is affected by this reaction has been proposed as a thermometer to study the peak temperatures of novae.  Previously, the $^{33}$S($p,\gamma$)$^{34}$Cl reaction rate had only been studied directly down to resonance energies of 432 keV.  However, for nova peak temperatures of $0.2-0.4$~GK there are 7 known states in $^{34}$Cl both below the 432 keV resonance and within the Gamow window that could play a dominant role.  Direct measurements of the resonance strengths of these states were performed using the DRAGON recoil separator at TRIUMF.  Additionally two new states within this energy region are reported.  Several hydrodynamic simulations have been performed, using all available experimental information for the $^{33}$S($p,\gamma$)$^{34}$Cl rate, to explore the impact of the remaining uncertainty in this rate on nucleosynthesis in nova explosions. These calculations give a range of  $\approx 20-150$ for the expected $^{33}$S overproduction factor, and a range of $\approx 100-450$ for the $^{32}$S/$^{33}$S ratio expected in ONe novae.
\end{abstract}



\maketitle

\section{Introduction}

Classical novae are thermonuclear explosions of the envelopes of white dwarf stars in accreting binary systems.  They occur when material from the companion star accreted onto the white dwarf is compressed in semi-degenerate conditions and a thermonuclear runaway occurs \cite{ClassicalNovae}.  This causes a dramatic increase in temperature, with peak luminosities reaching $\ge 10^4$~L$_{\footnotesize\astrosun}$.  The explosion also results in the ejection of $10^{-4} - 10^{-5}$~M$_{\footnotesize\astrosun}$ of material from the surface of the star, contributing to the chemical enrichment of the interstellar medium.  Observations of the chemical and isotopic abundances in the ejected shells can be used to test nova model predictions \cite{Jose2006}.

The $^{33}$S($p,\gamma$)$^{34}$Cl reaction is of particular importance in the study of oxygen-neon (ONe) novae as it affects two potential isotopic observables: $^{33}$S and $^{34m}$Cl.  $^{33}$S has the potential to be an important isotope for the classification of presolar grains \cite{Jose2004} and $^{34m}$Cl has been proposed as a potential target for $\gamma$-ray telescopes~\cite{Leising1987, Coc2000}.   Due to the short half-life of $^{34m}$Cl (31.99(3)~min~\cite{DataSheets}) compared to the time required for optical discovery of novae ($\sim 1-2$~weeks after reaching the peak temperature), $^{33}$S identification in presolar grains looks to be the most promising candidate.  Additionally, recent work by Downen~{\it et~al.}~\cite{Downen2013} has shown that the total S elemental abundance observed in the ejecta of a nova can be used as a thermometer to determine the peak temperature of the explosion.  This in turn provides a means of determining the mass of the underlying white dwarf.  While the uncertainty in the $^{30}$P($p,\gamma)^{31}$S reaction dominates the uncertainty in the total S abundance, this work reduces the uncertainty contribution from the $^{33}$S($p,\gamma$)$^{34}$Cl reaction.

A large overproduction of $^{33}$S compared to solar abundances has been predicted both in studies using rates determined from the Hauser-Feshbach statistical model~\cite{Iliadis2002} as well as studies by Jos\'{e}~{\it et~al.}~\cite{Jose2001} that used the available experimental data.    The result of the latter study was a calculated factor of 150 overproduction of $^{33}$S compared to solar. However, this may vary between $\approx1.5 - 450$ due to estimated experimental uncertainties \cite{Parikh2009}.   Data for the $^{33}$S($p,\gamma$)$^{34}$Cl reaction was taken from a complilation by Endt~\cite{Endt1990} and included resonance strength measurements by Waanders~{\it et~al.}~\cite{Waanders1983} down to a resonance energy of $E_r= 432$~keV.  

More recent work by Parikh~{\it et~al.}~\cite{Parikh2009} reports the existence of 6 new states in $^{34}$Cl located within the Gamow window for ONe nova peak temperatures (between 0.2~-~0.4~GK,  $E_r$~=~185~-~555~keV).  These states have the potential to significantly increase the $^{33}$S($p,\gamma$)$^{34}$Cl  rate, depending on their resonance strengths.  Based on a recommended rate, which is the geometric mean of the rate from only the known resonance strengths and the maximal theoretical contribution from the states with no resonance strength information, a factor of 12 reduction of the $^{33}$S abundance from the Jos\'{e}~{\it et~al.}~\cite{Jose2001} value was estimated~\cite{Parikh2009}.  
The goal of this work is to measure resonance strengths or at least determine sufficiently restrictive upper-limits of the as yet unmeasured states relevant to ONe novae nucleosynthesis.


If $^{33}$S proves to have a significant signature in presolar grains it would be a particularly useful diagnostic for characterizing grains of nova origin as a large overproduction factor would not be expected from supernova nucleosynthesis scenarios~\cite{Jose2004}.  There remain a few technical challenges, however, as measurements of S isotopes are complicated by possible contamination introduced during the chemical separation of these grains from the surrounding medium~\cite{Amari2001}.  Fortunately, there has been significant progress measuring S ratios in presolar grains from other astrophysical environments in the past few years~\cite{Hoppe2010,Zinner2010,Hoppe2012,Hoppe2012LPI,Gyngard2012}.  This makes the measurements of resonance strengths of the $^{33}$S($p,\gamma$)$^{34}$Cl resonances presented in this work particularly timely.

\section{Experimental}

This work was performed using the DRAGON (Detector of Recoils and Gammas Of Nuclear reactions) recoil separator located at the ISAC radioactive beam facility at the TRIUMF laboratory in Vancouver, Canada.  A beam of $^{33}$S$^{6^+}$ was produced using an enriched sulphur sample placed in the oven of a Supernanogan ECR source, part of the ISAC Off-line Ion Source (OLIS).  The beam was accelerated to energies between $194 - 514$~keV/nucleon and impinged on the DRAGON windowless gas target~\cite{Hutcheon2003}.  The average beam intensity was $1.16\times 10^{10}$~s$^{-1}$. The target consisted of $7.3-8.1$~mbar of H$_2$ gas with pressures regulated to 0.04~mbar or better during each hour-long run.  These pressures resulted in energy losses across the target of $\sim16-20$~keV/nucleon.

The gas target volume was surrounded by 30 Bismuth Germanate (BGO) detectors which detected the $\gamma$~rays emitted during the reaction~\cite{Hutcheon2003}.  In addition to providing information about the number and energy of these $\gamma$~rays, the application of a coincidence requirement ($\le10~\mu$s separation) between  $\gamma$~rays and the heavy ion events detected at the end of the separator was used to increase the beam suppression~\cite{Hutcheon2008}.  The segmented nature of the BGO~array also provided a means of determining where within the target volume the reactions took place.  This information could then be used to determine the resonance energy of the reaction~\cite{Hutcheon2012}.

Downstream of the target, the reaction products were separated from the unreacted beam by a recoil separator consisting of two magnetic (M) and two electrostatic (E) dipoles, arranged in an MEME configuration.  Charge state, $q$, selection was performed after the first magnetic dipole and $A/q$ selection was performed after the first electrostatic dipole.  The second stage of separation functioned in the same way as the first and provided the additional beam suppression required for the typically low reaction rates of the reactions of interest~\cite{Hutcheon2003,Hutcheon2008,Engle2005}.

The recoils transmitted through the separator were detected in a double sided silicon strip detector (DSSSD)~\cite{Wrede2003}.  Additionally a local time-of-flight (TOF) system, consisting of two microchannel plate (MCP) detectors~\cite{Vockenhuber2009}, was used to increase the beam suppression through improved particle identification.

\section{Data and Analysis}

\begin{table}
\caption{\label{tab:resonances} Known $^{33}$S$+p$ resonance energies, $E_r$, derived from their corresponding $^{34}$Cl excitation energies, $E_x$, and the $Q$ value of this reaction $5143.21(5)$~keV \cite{AME12} listed along with the energy range covered by the target.  The uncertainty on $E_r$ is the same as $E_x$ in all cases below.}
\begin{center}
\begin{tabular}{llcc}
\hline
\multicolumn{1}{c}{ $E_r$	}&\multicolumn{1}{c}{  $E_x$}&\multicolumn{1}{c}{ $E_{\rm c.m.}$ range }& measurement no.~\\
\multicolumn{1}{c}{  [keV]}&\multicolumn{1}{c}{ [keV] }&\multicolumn{1}{c}{ [keV]} &  \\
 \hline\hline
491.8   &   5635.0(5)$^{\ddag}$&   $483-503$	& 1\\
\hline
432   &   5575(2)$^{\ddag}$&   $426-443$	& 2\\
	&	&	$425-441$	& 3\\
	&	&	$424 - 443$	& 4\\
\hline
399  &     5542(2)$^{\ddag}$& $394-412$	& 5\\
\hline
342   &   5485(4)$^{*}$&   $336-353$ 	& 6\\
\hline
&   &   $303-321$	& 7\\
\hline
301   &   5444(4)$^{*}$&   $295-312$	& 8\\
	&	&	$292-309$	& 9\\
	&	&	$287-304$	& 10\\
\hline
281   &   5424(4)$^{*}$&   $275-291$  	& 11\\
\hline
243.6   &   5386.8(15)$^{\dag}$&  $238-254$   	& 12\\
\hline
214   &   5357(4)$^{*}$&  $208-223$	&	13\\
	&	&	$204-220$   & 14\\
\hline
183   &   5326(4)$^{*}$&  $176-190$   	& 15\\				
 \hline
\multicolumn{3}{l}{\footnotesize $^{*}$ Parikh~{\it et~al.}~\cite{Parikh2009}}\\
\multicolumn{3}{l}{\footnotesize $^{\dag}$ Endt~\cite{Endt1990}}\\
\multicolumn{3}{l}{\footnotesize$^{\ddag}$ weighted average of \cite{Parikh2009} and \cite{Endt1990}}
\end{tabular}
\end{center}
\end{table}

Measurements were performed for 9 known resonances corresponding to states in $^{34}$Cl located within the Gamow window range of  peak temperatures in ONe novae. These measurements (Table \ref{tab:resonances}) will be referenced in the text below according to the corresponding measurement number.  Beam energies were chosen to place the resonance in the centre of the target.  The exception to this is in the region of centre of mass energies, $E_{c.m.}=275 - 321$~keV (measurements 7, 8, 9, 10 and 11) where a range of beam energies was used to better study the level structure in this region.

To extract a resonance strength, $\omega\gamma$, from these measurements we first needed to determine the yield, $Y$, from the number of detected recoil events, $n_{\rm recoils}$, and the total number of beam particles, $N_{\rm beam}$, given all of the various separator and detector efficiencies:

	\begin{equation}
	\label{eq:yield}
	Y=\frac{n_{\rm recoils}}{N_{\rm beam}~\eta _{\rm BGO} \eta _{\rm sep} \eta _{\rm CSF} \eta _{\rm MCP_d}  \eta _{\rm MCP_t} \eta_{\rm DSSSD}  \eta_{\rm live}}.
	\end{equation}

\noindent These efficiencies are: the BGO $\gamma$ detection efficiency, $\eta _{\rm BGO}$; the separator transmission, $\eta _{\rm sep}$; the charge state fraction for the selected recoil charge state, $\eta _{\rm CSF}$; the MCP detection, $\eta _{\rm MCP_d}$, and transmission efficiency, $\eta _{\rm MCP_t}$; the DSSSD detection efficiency, $\eta _{\rm DSSSD}$; and the live-time of the data acquisition system, $\eta _{\rm live}$.

The resonance strength can then be calculated as follows:

	\begin{equation}
	\label{eq:strength}
	\omega\gamma = \frac{2 \epsilon Y}{\lambda^2}\frac{M_{\rm target}}{M_{\rm beam}+M_{\rm target}}
	\end{equation}

\noindent where $\epsilon$ is the stopping power of the target, $\lambda$ the deBroglie wavelength and $M_{\rm beam}$ and $M_{\rm target}$ the beam and target masses, respectively.

\subsection{$^{34}$Cl identification}

\begin{figure}[tb]
 \begin{center}
	 \includegraphics[width=8.5cm]{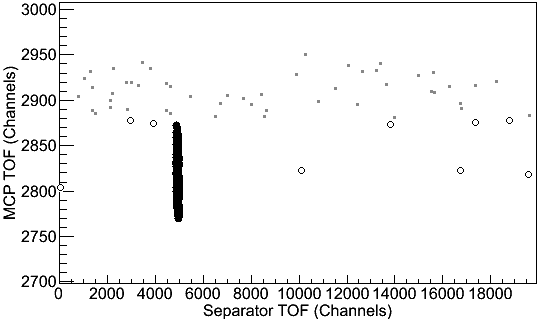}
\end{center}
\caption{\label{fig:SepTOFMCPTOF1}Plot of MCP TOF vs.~separator TOF for data passing DSSSD energy cuts for measurement no.~4, $E_r$ = 432 keV (see Table~\ref{tab:resonances}).  Grey squares represent coincidence events, solid black circles indicate events which pass both MCP and separator TOF cuts and are the recoils from the reaction, and the open circles indicate events in the background region (within the MCP TOF cut, but outside the separator TOF cut)}
 \end{figure}

\begin{figure}[tb]
	 \begin{center}	 
	 \includegraphics[width=8.4cm]{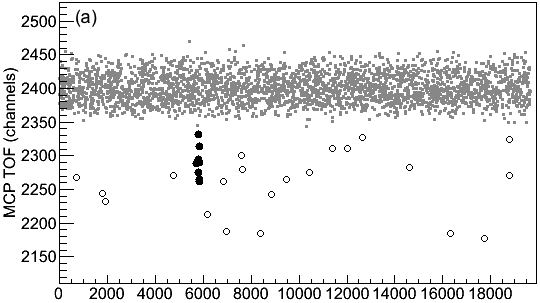}
	 \includegraphics[width=8.4cm]{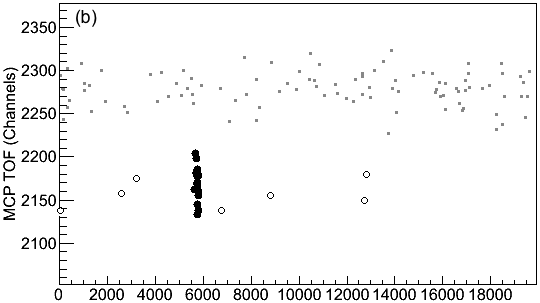}
	  \includegraphics[width=8.4cm]{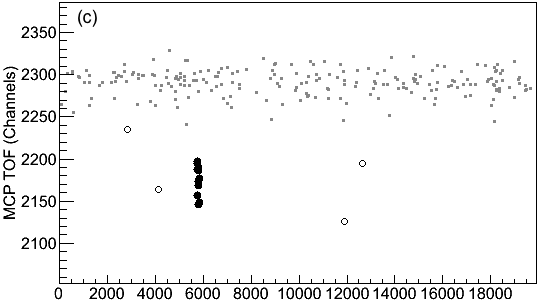}
	  \includegraphics[width=8.4cm]{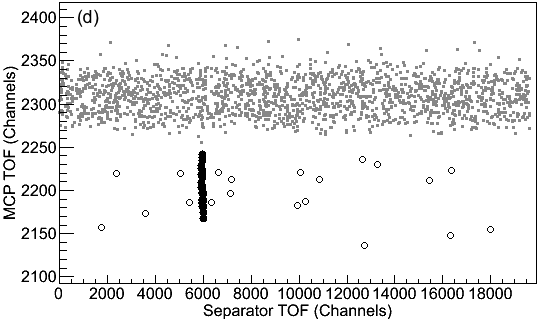}
	 \end{center}
 \caption{\label{fig:SepTOFMCPTOF2} Same as Fig. \ref{fig:SepTOFMCPTOF1} but for measurements $7 - 10$ (Table~\ref{tab:resonances}).}
 \end{figure}
 
The individual particle identification of the $^{34}$Cl reaction products (recoils) and unsuppressed, unreacted $^{33}$S beam particles (``leaky beam") was performed by placing cuts on both the local TOF between the two MCP detectors (MCP TOF), and the total TOF between the detection of coincident $\gamma$ rays and heavy ions (separator TOF).  The separator TOF varied between $2.2$ and $3.5~\mu$s depending on the incoming beam energy whereas the MCP TOF ranged from $60$ to $100$ ns.



\begin{figure*}[tb]
	  \includegraphics[width=8.5cm]{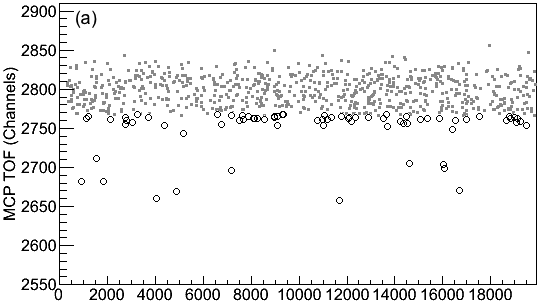}
	   \includegraphics[width=8.5cm]{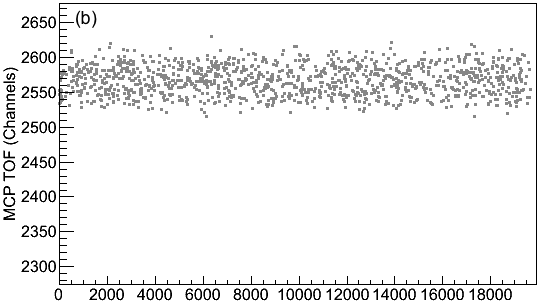}
	    \includegraphics[width=8.5cm]{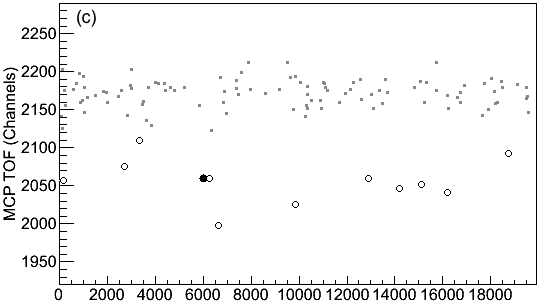} 
	      \includegraphics[width=8.5cm]{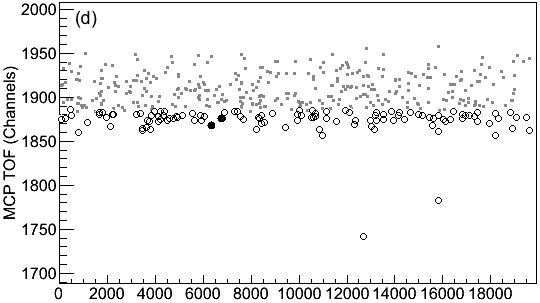}
		 \includegraphics[width=8.5cm]{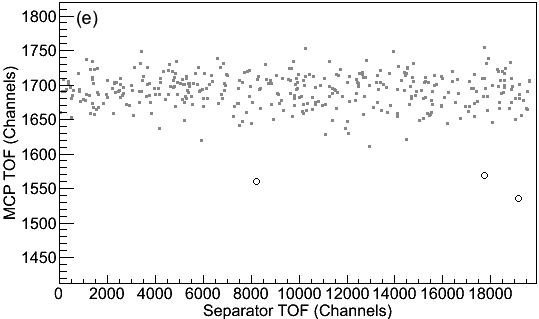}
		  \includegraphics[width=8.5cm]{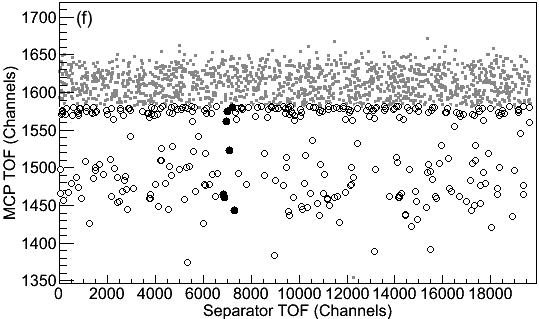}
		    \includegraphics[width=8.5cm]{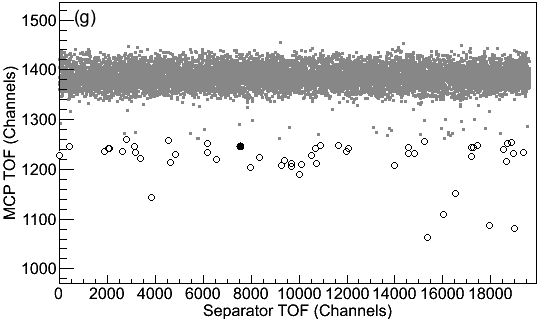} 
	\caption{\label{fig:SepTOFMCPTOF3} Same as Fig. \ref{fig:SepTOFMCPTOF1} but for all measurements consistent with no recoils, for which only upper limits could be determined.  From left to right, top to bottom: measurements 5, 6, 11, 12, 13, 14 and 15 (see Table~\ref{tab:resonances}).}
 \end{figure*}

Plots of the data showing the MCP TOF versus the separator TOF and the effects of the various TOF cuts are presented in Figures \ref{fig:SepTOFMCPTOF1}, \ref{fig:SepTOFMCPTOF2}, and \ref{fig:SepTOFMCPTOF3}.  The number of events passing the cuts (recoils) and the expected background (calculated by determining the number of events/channel from the background region within the MCP TOF cut, but outside the separator TOF) are presented in Table \ref{tab:recoils}.

\subsection{Beam normalization}

Silicon detectors  (SD), located 30$^{\circ}$ and 57$^{\circ}$ from the beam axis inside the gas target, were used to detect the elastic scattering of H nuclei by the beam as a means of continuously monitoring the beam intensity.  Faraday cup readings of the beam current ($I$) taken before and after each run were matched with the SD scaler rate ($N_H/\Delta t$) of the first and last few minutes of each run to determine the normalization factor, $R$:

\begin{equation}
	\label{eq:beam_norm}
	R=\frac{I P}{e q_{\rm b}}\frac{\Delta t}{N_{\rm H}},
	\end{equation}

\noindent where $q_{\rm b}$ is the charge state of the incoming beam, $P$ the H$_2$ gas pressure in Torr and $e$ the proton charge. $N_H$ indicates the number of elastically scattered H nuclei detected.

This technique was used for the majority of measurements.  There were, however, a few runs where the SD threshold did not exclude a low energy peak in the SD energy spectrum.  While this peak was determined to be beam related, the ratio of the low energy peak to the elastically scattered proton peak was not constant across beam intensities, meaning that the total SD scaler rate could not be used for normalization in these cases.  In these cases a modified procedure was used.  Instead of using the shorter portions of the SD scaler events to calculate $R$, the total number of SD events in the scattered proton peak over the whole run was used.  This placed a stringent requirement on which runs could be used for this calculation:  runs were limited to cases where the beam had a constant SD scaler rate with no interruptions, where Faraday cup readings were performed both immediately before and after the run, and where these readings agreed within uncertainties.  Sufficient runs meeting these criteria were available for each energy for which this was necessary.  $R$ was again calculated using equation \ref{eq:beam_norm} but with $N_H$ representing the total number of SD events in the proton peak, $\Delta t$ representing the total duration of the run in seconds, and $I$ representing the average of Faraday cup readings bracketing the run.

 The total number of incident ions in each run can be then calculated from:
 	\begin{equation}
	\label{eq:ninc}
	N_{\rm beam}=\frac{N_{\rm Htot} R}{P},
	\end{equation}
where $N_{\rm Htot}$ is either the total number of SD scaler events, or number of SD events within the proton scattering peak, depending on the case.  The resulting $N_{\rm beam}$ is given in Table \ref{tab:recoils}.  The uncertainty in $N_{\rm beam}$ is dominated by the uncertainty in $R$.

\begin{table}
\caption{\label{tab:recoils} The number of events passing the cuts (recoils) and the expected background (see text) for each measurement.  The  $E_{\rm c.m.}$  range covered by the target is listed. }
\begin{center}
\begin{tabular}{ccccc}
\hline
 $E_{\rm c.m.}$ range &~recoils~& expected&$N_{\rm beam}$&$\Delta N_{\rm beam}$ \\
~[keV]& & ~background~ & &$\%$ \\
\hline
$483-503$	&~608	&0.4&$2.28\times10^{13}$&	3.2	\\
$426-443$	&~489	&0.0	&$1.87\times10^{13}$&	3.3	\\
$425-441$	&1664	&0.7	&$7.25\times10^{13}$&	3.1	\\
$424-443$	&~976	&1.4	&$4.26\times10^{13}$&	3.1	 \\
$394-412$	&~~~0		&1.6	&$6.16\times10^{14}$&	3.5	\\
$336-353$	&~~~0		&0.0	&$3.23\times10^{14}$&	4.5	\\
$303-321$	&~~~8		&0.5	&$8.66\times10^{14}$&	6.5	\\
$295-312$	&~~12		&0.1	&$5.49\times10^{14}$&	3.3		\\
$292-309$	&~~20		&0.2	&$1.24\times10^{15}$&	8.2	\\
$287-304$	&~~26		&0.5	&$6.07\times10^{14}$&	7.2	\\
$275-291$	&~~~1		&0.3	&$1.03\times10^{15}$&	9.9		\\
$238-254$  	&~~~2		&2.8	&$5.24\times10^{14}$&	8.6	\\
$208-223$	&~~~0		&0.1	&$1.11\times10^{14}$&	5.9		\\
$204-220$	&~~~7		&7.9	&$3.97\times10^{15}$&	7.3		\\
$176-190$	&~~~1		&1.4	&$1.47\times10^{15}$&	11.6		\\
 \hline

\end{tabular}
\end{center}
\end{table}

\subsection{Recoil charge state distributions}

The passage of the beam and recoils through the gas target results in a distribution of charge states which is dependant on the atomic number of the ion and its velocity.  As the recoil separator can only accept a single charge state, this distribution needs to be accounted for.

Measurements were performed using a $^{35}$Cl beam at incoming beam energies chosen such that the speed upon exiting the target matched that of a recoil during the $^{33}$S($p,\gamma)^{34}$Cl experiment. The beam current was measured in a Faraday cup located after the charge state selection slits (FCCH), for each $q$ accessible within the magnetic field range of the magnetic dipole.  To account for and normalize against any fluctuations in beam intensity, measurements of the incoming beam intensity were taken on a Faraday cup upstream of the target both before and after the FCCH measurement.

Five energies were measured, corresponding to measurements no.~1, 5, 7, 14 and 15 (Table \ref{tab:csd}).  The other charge state fractions were interpolated using this data and equations 14 and 15 from \cite{Liu2003}.

\begin{table}
\caption{\label{tab:csd} Charge state fractions of each measured charge state for the five  $^{35}$Cl beam energies studied. E$_{\rm beam\_out}$ is the energy in keV/nucleon after passing through the $\approx 8$ mbar of H$_2$ gas in the target.}
\begin{center}
\begin{tabular}{ccl}
\hline 
 E$_{\rm beam\_out}$&$~~q~~$&\multicolumn{1}{c}{CSF (\%)}\\
\hline 
467.6(9) 	&	7	&	45.8	(2.2)\\
		&	8	&	28.3	(1.3)\\
		&	9	&	~6.9	(0.8)\\
\hline 
375.3(8)	&	6	&	35.3	(1.0)\\
		&	7	&	29.3	(0.8)\\
		&	8	&	~5.8	(0.3)\\
\hline 
294.1(6)	&	5	&	36.2	(1.0)\\
		&	6	&	26.9	(0.8)\\
		&	7	&	~7.5	(0.4)\\
		&	8	&	~0.7	(0.2)\\
\hline
199.1(4)	&	5	&	27.0	(0.7)\\
		&	6	&	~4.9	(0.6)\\
\hline
172.5(3)	&	4	&	44.8	(2.6)\\
		&	5	&	18.4	(1.7)\\
		&	6	&	~3.3	(1.1)\\
\hline 
\end{tabular}
\end{center}
\end{table}

\subsection{Separator transmission}

\begin{table}
\caption{\label{tab:eff} Table of the efficiencies (in \%) that either change as a function of energy or vary from measurement to measurement, listed for each $E_{\rm c.m.}$ range studied (keV). The charge state of the chosen recoils is also listed.}
\begin{center}
\begin{tabular}{lccllcc}
\hline
$E_{\rm c.m.}$  range  &	$~q~$	& $\eta_{\rm CSF}$ & \multicolumn{1}{c}{$\eta_{\rm BGO}$} & \multicolumn{1}{c}{$\eta_{\rm sep}$} & $\eta_{\rm MCP_d}$&$\eta_{\rm live}$ \\
\hline
$483-503$	&7	&46(2)	&69(10)$^*$	&95.6(5)$^*$	&99(1)&95(1)	\\
$426-443$	&7	&38(3)	&66(10)$^*$	&96.7(5)$^*$	&99(1)&97(1)	\\
$425-441$	&7	&38(3)	&69(10)$^*$	&96.7(5)$^*$	&88(1)&99(1)	\\
$424 - 443$	&7	&37(3)	&69(10)$^*$	&96.7(5)$^*$	&98(1)&98(1)	\\
$394-412$	&6	&29(1)	&77(10)$^{\dag}$	&98.8(5)$^{\dag}$	&90(1)&99(1)	\\
$336-353$	&6	&32(3)	&79(22)	&99.3(5)	&95(5)&99(1)		\\
$303-321$	&6	&27(1)	&78($^{+13}_{-27})$	&99(1)	&87(1)&98(1)		\\
$295-312$	&6	&25(1)	&74($^{+10}_{-19})$	&99.1(5)	&99(1)&98(1)		\\
$292-309$	&6	&24(1)	&74($^{+10}_{-19})$	&99(1)	&99(1)&99(1)		\\
$287-304$	&6	&23(1)	&79($^{+13}_{-27})$	&99(1)	&90(1)&98(1)		\\
$275-291$	&6	&20(1)	&74(19)	&98.9(5)	&99(1)&98(1)		\\
$238-254$	&6	&11(1)	&77(10)$^{\dag}$	&98.1(5)$^{\dag}$	&99(1)&98(3)		\\
$208-223$	&5	&27(1)	&73(19)	&98.9(5)	&99(1)&78(1)		\\
$204-220$	&6	&5(1)	&77(21)	&98.9(5)	&80(1)&99(2)		\\
$176-190$	&4	&45(3)	&78(20)	&98.4(5)	&86(1)&98(1)		\\
 \hline
 \multicolumn{7}{l}{\footnotesize $^*$ Determined using branching ratios from Freeman~{\it et al.}~\cite{Freeman2011}}\\
 \multicolumn{7}{l}{\footnotesize $^{\dag}$ Determined using branching ratios from Endt~\cite{Endt1990}}

\end{tabular}
\end{center}
\end{table}

The recoil cone angle, and thus the separator transmission, depends on both the energy of the beam and recoils as well as the $\gamma$ cascade of the $^{34}$Cl state populated in the reaction.  Therefore, the separator transmission was determined independently for each measurement.  For states where the branching ratios were known (see Table \ref{tab:eff}) this information was included in the GEANT3 simulation used to determine the separator transmission.  For states where the $\gamma$ cascade was not known the transmission was determined using GEANT3 simulations for cascades of 1, 2, 3 or 4 $\gamma$ rays.  As the recoil cone angle for the $^{33}$S($p,\gamma)^{34}$Cl reaction is small compared to the DRAGON acceptance at these energies the transmission for all possible $\gamma$ cascades agreed within the statistical uncertainty of the simulations.

The separator transmission could also have been affected by the width of the resonance and its position in the target. While the beam energy was chosen to place the resonance in the central portion of the target, there was still some uncertainty in our energy determinations as well as in the expected resonance energy itself. To account for this systematic uncertainty, GEANT3 simulations were run for cases where the resonance was positioned at varying intervals from the centre of the target.  As long as the resonance was within the central $\pm4$~cm of the physical target length (i.e. the central 73\%), the transmission agrees with the values given in Table \ref{tab:eff}.  This more than accounts for any position variation based on the 0.2\% uncertainty in the determination of the beam energy~\cite{DAuria2004} and the uncertainty in the resonance energy (Table \ref{tab:resonances}); the largest combined uncertainty, when coupled with the smallest stopping power of the target, would give a position uncertainty of $\pm2.3$~cm.  Also, from GEANT simulations it was determined that as long as the resonance was in the central portion of the gas target and narrow enough to be contained within the target volume, the variation in transmission was still within the uncertainty listed in Table \ref{tab:eff}.

For measurements no.~7, 9 and 10 it is possible that the resonance was located at the upstream side of the target, outside of this $\pm4$~cm region (see section \ref{sect:disc}). In these cases the uncertainty on the transmission is larger to account for the additional loss of transmission on the upstream portions of the gas target.  Regardless of position, this uncertainty is only a small contribution to the total error budget. 

\begin{figure*}[tb]
	 \begin{center}
	 \includegraphics[width=12cm]{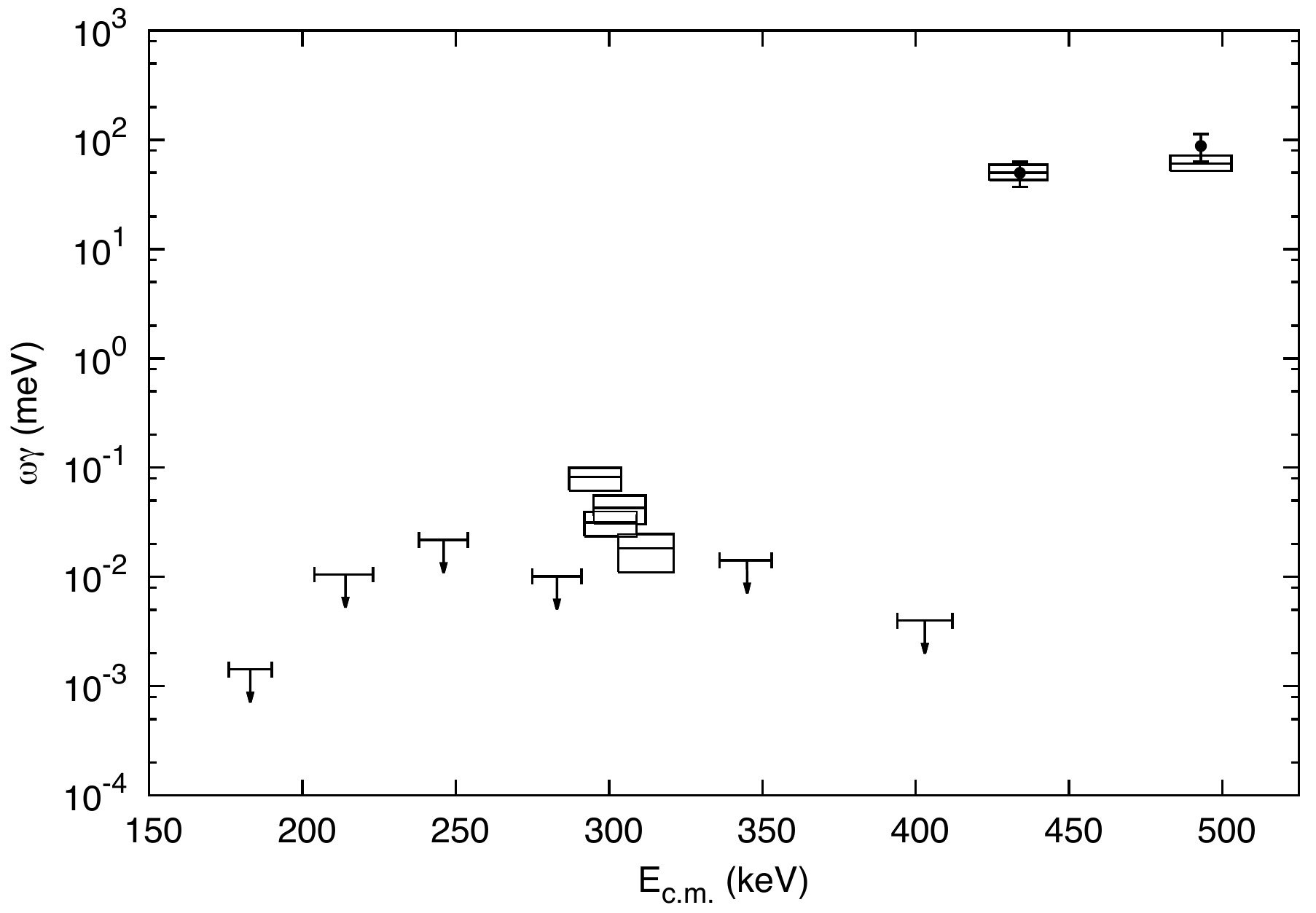}
	 \end{center}
	 \caption{\label{fig:wg} $\omega\gamma$ for each $E_{\rm c.m.}$ range measured. The upper limits (95\% confidence level) are plotted as an $E_{\rm c.m.}$ error bar to indicate the energy range covered by the target, with a downwards arrow.  The measured $\omega\gamma$ are indicated by boxes covering both the target energy range as well as the uncertainty on $\omega\gamma$ (see section \ref{wgerr}). Where multiple measurements at the same energy were performed, they have been combined using a weighted average. Literature values from \cite{Endt1990} are indicated by the black circular data points. Note that for measured $\omega\gamma$ the $E_r$ value can be determined more precisely within the energy range covered by the target (see Section \ref{sec:Er}).   }
 \end{figure*}

\subsection{DSSSD and MCP detection efficiency}

The detection efficiency of the DSSSD and the MCP transmission efficiency were both constant across all measurements.  These values have been determined previously and were 97.0(7)\% \cite{Wrede2003} and 76.9(6)\% \cite{Vockenhuber2009}, respectively.

The MCP detection efficiency depends on the energy of the leaky beam and recoil particles as well as on the MCP detector threshold which was occasionally adjusted.  This efficiency was determined either from recoil events in cases where they were plentiful (measurements no.~1, 2, 3, and 4) or from the attenuated beam run at each beam energy.  There was only one case where no attenuated beam run was performed and where there were insufficient recoils to determine this value.  The MCP detection efficiency used in this case was the average of all MCP detection efficiencies with the standard deviation as the uncertainty (Table \ref{tab:eff}).

\subsection{BGO efficiency}

Knowledge of the $\gamma$ branching ratios of the  $^{34}$Cl states populated in these reactions is particularly helpful when determining the BGO detector efficiency as it depends on both the number of  $\gamma$ rays emitted and their energies.  As a result this efficiency is determined on a case-by-case basis through GEANT3 Monte Carlo simulations \cite{Dario}.

The branching ratios are known for 4 of the states listed in Table \ref{tab:resonances}.  Values from the recent work by Freeman~{\it et\,al.} \cite{Freeman2011} were used where available ($E_x$ = 5635 and 5575 keV).  For the other states, $E_x$ = 5542 and 5387 keV, the branching ratios were taken from \cite{Endt1990}.  The resulting efficiencies are listed in Table \ref{tab:eff}.

For the measurements where the branching ratios are not known, the BGO efficiencies were determined by running simulations where the total energy of the state was  split evenly among 1, 2, 3 or 4 $\gamma$ rays. As all of the nearby known states decay preferentially via 2 $\gamma$ rays above the BGO threshold, the 2 $\gamma$-ray case was taken as the BGO efficiency, with the difference in efficiency between it and the 1 and 3 $\gamma$-ray cases combined with the 10\% standard uncertainty in the efficiency.  In many cases, the 2 $\gamma$-ray case gave the maximum BGO efficiency; in those cases the positive uncertainty was taken to be the standard 10\% uncertainty.  These efficiencies are also listed in Table \ref{tab:eff}.  For measurements where upper limits on $\omega\gamma$ were determined only the negative uncertainty is listed as the lower efficiency would result in a higher $Y$, and therefore $\omega\gamma$.  

For states with unknown branching ratios, the uncertainty on the BGO efficiency can be up to 3 times larger than the standard uncertainty (Table\ref{tab:eff}).  Measurements of these branching ratios are encouraged as they could potentially reduce the upper limits reported in this work.

\subsection {Calculating upper limits and confidence intervals on $Y$}\label{wgerr}

The upper limits and confidence intervals were calculated using Ref.~\cite{Rolke2005}, which details the calculation of ``Limits and confidence intervals in the presence of nuisance parameters" including statistical and systematic uncertainties on the efficiency values.  In this work the uncertainty on $N_{\rm beam}$ was also included in this uncertainty.

For consistency this was used to determine the uncertainties or upper limits for all of the data.  For the measurements, the 68\% confidence intervals define the error bars, whereas the upper limits given are the 95\% confidence limits.  The only modification to this procedure was in the case of measurements no.~$7 - 10$.  In these cases the uncertainty on the BGO efficiency did not fit the Gaussian or Poisson criteria for the uncertainty on the efficiencies.  In these cases the BGO uncertainty was added in quadrature with the resulting 68\% error bars from the confidence interval.  Note that this could result in an overestimation of the uncertainty.

The resulting $\omega\gamma$ values and upper limits are plotted in Figure \ref{fig:wg}.   Of the 10 $\omega\gamma$ values presented only two, $E_r = 492$ and 432~keV,  have been previously measured.  Both values from this work are in good agreement with those measured by Waanders~{\it et~al.}~\cite{Waanders1983}.

\subsection{Resonance energy determinations}\label{sec:Er}

\begin{figure}[tb]
	 \begin{center}
	 \includegraphics[width=9cm]{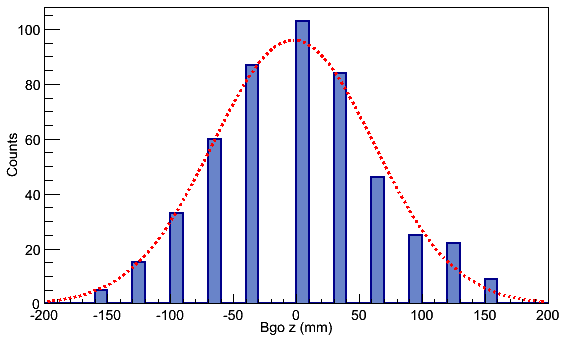}
	 \end{center}
	 \caption{\label{fig:BGOposCenter} (Color online) The BGO position spectra for measurement no.~2 (see Table~\ref{tab:resonances}). Note that as the x axis indicates the position along the beam axis, higher numbers correspond to lower beam energies.}
 \end{figure}

 For all the upper limit determinations, the $E_r$ value from the literature (Table \ref{tab:resonances}) was adopted.  For measurements where coincident recoil events were measured, we used the position information of the BGO detector in which the $\gamma$ rays were detected to determine the location of the resonance in the target.  An example BGO position distribution (measurement no.~2) is presented in Fig. \ref{fig:BGOposCenter}.  When this information was combined with the beam energy and energy loss information, it provided a means of determining $E_r$ \cite{Hutcheon2012}.  The precision with which $E_r$  could be determined depended on both the inherent resolution of the BGO array as well as the number of detected events.  With sufficient statistics, $E_r$ may be determined to a precision of 0.5\%.

The procedure laid out in \cite{Hutcheon2012} was followed for measurements 1, 2, 3 and 4.  The resulting $E_r$ values for measurements 2, 3, and 4 were then combined via weighted average. The $E_r$ determinations for these two resonances are given in Table \ref{tab:wg}.  

\begin{figure}[tb]
	 \begin{center}
	  \includegraphics[width=9cm]{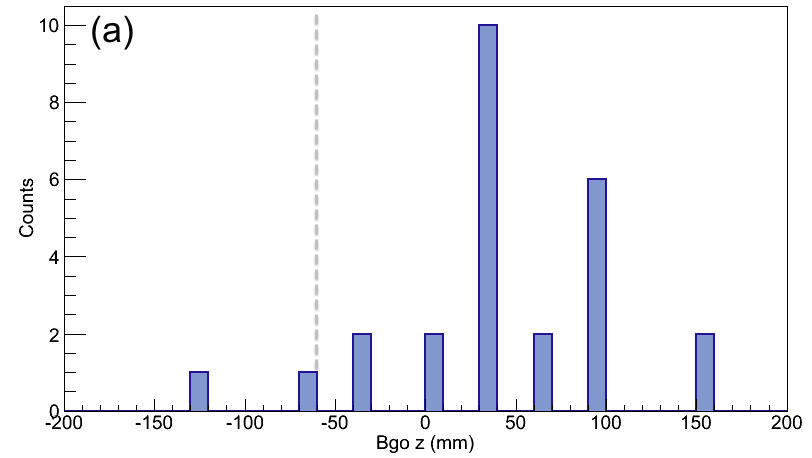}
	 \includegraphics[width=9cm]{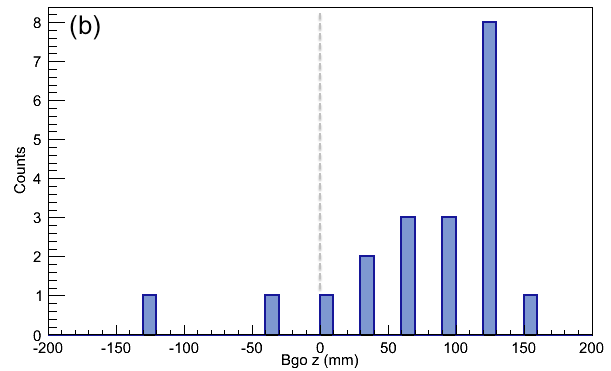}
	 \includegraphics[width=9cm]{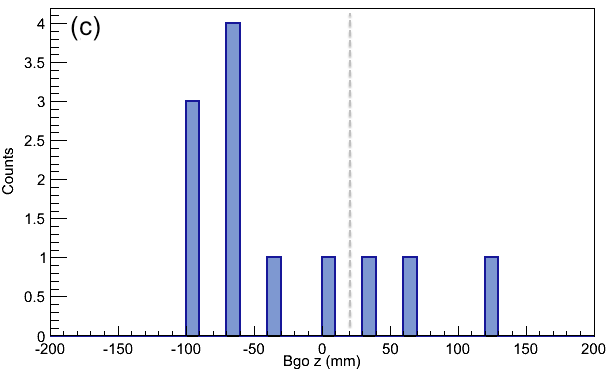}
	
	 \end{center}
	 \caption{\label{fig:BGOpos} (Color online) The BGO position spectra for measurements no.~a)~10, b)~9 and c) 8 (see Table~\ref{tab:resonances}). The dashed grey line indicates the expected peak centroid for a resonance at $E_r$ = 301 keV (note that as the x axis indicates the position along the beam axis, higher numbers correspond to lower beam energies).}
 \end{figure}

None of the 3 measurements which cover the 301 keV resonance energy (no. 8, 9 and 10) show an indication of a peak at the expected BGO positions, and instead show peaks at either upstream or downstream positions (Fig~\ref{fig:BGOpos}).   We are unable to place a numerical upper limit on the $\omega\gamma$ of the 301 keV resonance, but we can say from this information that it will be a small fraction of the $\omega\gamma$ of measurements 8, 9 and 10 and will thus not be astrophysically significant.   Given that the measurement of the only known nearby resonance ($E_r = 281$~keV, measurement~no.~11) yielded an upper limit, one or more previously unknown resonances must be present to account for the yields seen in measurements 7, 8, 9 and 10.

Due to the low count rates and non-central position of the $\gamma$-ray distributions, the method used to determine $E_r$ values for the two highest energy resonances is not valid for measurements 7, 8, 9, and 10.  Some resonance energy information can still be gleaned from the BGO position information, though it involves a much more qualitative discussion.

\begin{figure}[tb]
	 \begin{center}
	 \includegraphics[width=9cm]{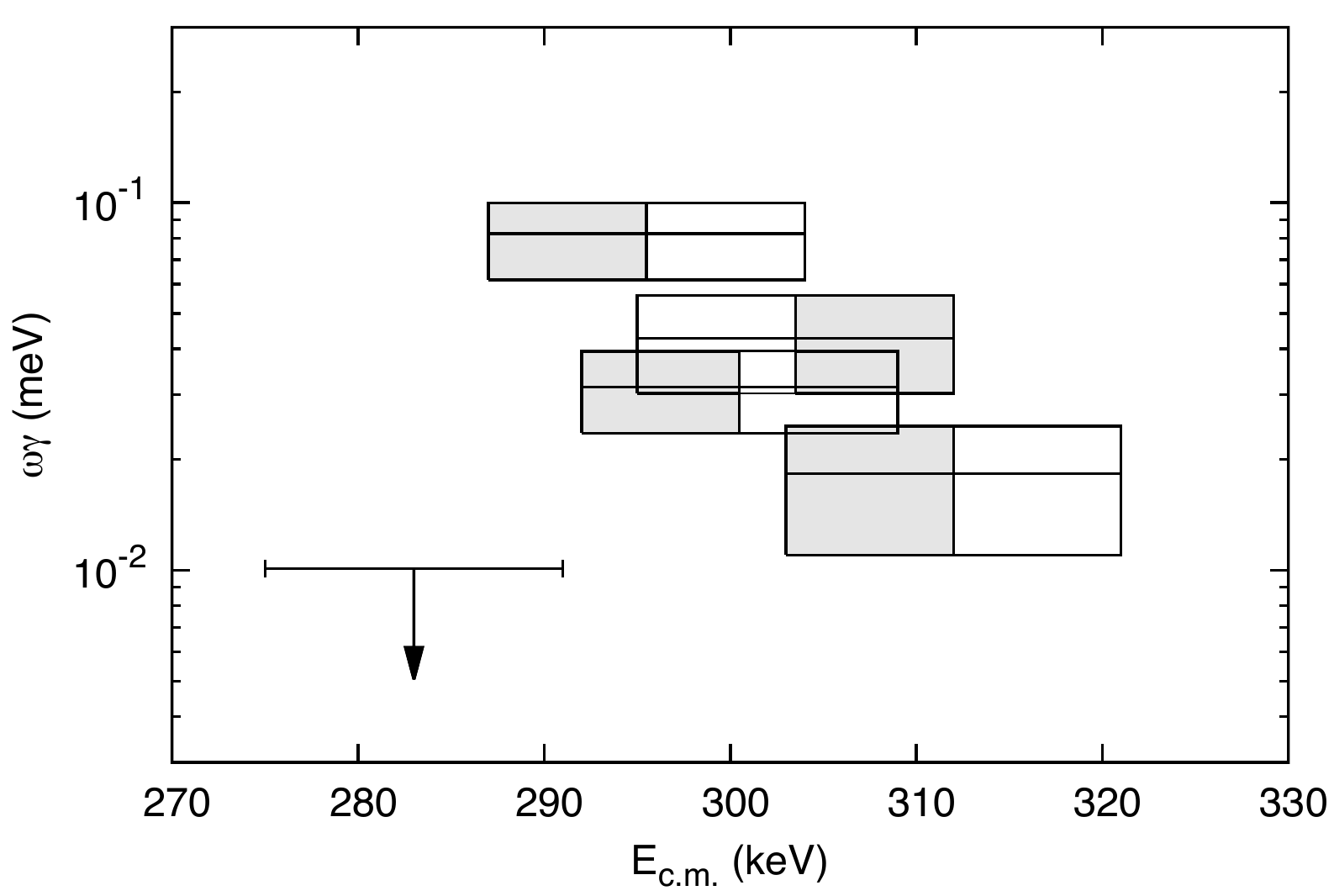}
	 \end{center}
	 \caption{\label{fig:BGOposEr} Energy range covered by the gas target vs. the measured resonance strength (or upper limit) for measurements 7, 8, 9, 10 and 11 (from right to left)(see Table~\ref{tab:resonances}). The shaded regions indicate whether the peak in the BGO position spectrum is located in the upstream or downstream portion of the target.}
 \end{figure}
 
If we limit the possible energy range for the resonance in these 4 measurements to the half of the target in which the BGO position peak is located, there is already an indication that there are two separate resonances (Fig.~\ref{fig:BGOposEr}).  We can then use the combination of measurements in this region to narrow down the possible $E_r$ values.  One telling feature is the dramatic change in the BGO position spectra between measurements 8 and 9, which differ by only 3 keV in incoming beam energy (Fig.~\ref{fig:BGOpos}).  The resonance present in the very upstream portion of the target in measurement~8 (Fig.~\ref{fig:BGOpos}c)  appears to be excluded from the target region in measurement~9 (Fig.~\ref{fig:BGOpos}b).  Similarly, the resonance at the very downstream end of the target in measurement~9 is not present in measurement~8.   When combined with the upper limit determined for measurement~11 the possible resonance energy for the lower energy of the two proposed resonances is $291-295$ keV.  Looking at the $\omega\gamma$ of the two measurements which included this state (9 and 10) it seems likely that the exit energy of measurement no.~9 ($E_{c.m.}=292$ keV/nucleon) lies on the slope of the thick target yield curve~\cite{IliadisText} and would therefore be close to the $E_r$ of this resonance.  Given the uncertainties, however, it cannot be determined where on the slope this measurement lies, and it cannot provide any further energy information.  The $\omega\gamma$ assigned to this potential resonance is that of measurement no. 10.
	Looking at the $\omega\gamma$ values of the two measurements that include the higher energy of the two proposed resonances (7~and~8) it appears that both would be on the plateau of the thick target yield curve.  If this is the case, then this resonance is entirely located in the 3 keV region covered in measurement~8, and excluded in measurement~9, in other words, within $309-312$ keV.  The $\omega\gamma$ assigned to this potential resonance is the weighted average of measurements 7 and 8.

\begin{table}
\caption{\label{tab:wg} $\omega\gamma$ and upper limits ($95\%$ C.L.) along with measured $E_r$ values, where possible.}
\begin{center}\begin{tabular}{llcccr}
\hline
 \multicolumn{3}{c}{literature}&~~& \multicolumn{2}{c}{this work}   \\
 \cline{1-3}\cline{5-6}
 \multicolumn{1}{c}{$E_r\,^*$~~} & \multicolumn{1}{c}{$\omega\gamma^{\dag}$}& \multicolumn{1}{c}{$(\omega\gamma)^{\ddag}$ }&~~& \multicolumn{1}{c}{$E_{\rm r}$}& \multicolumn{1}{c}{$\omega\gamma$}  \\
 \multicolumn{1}{c}{[keV]~~}& \multicolumn{1}{c}{measured}& \multicolumn{1}{c}{calc.~up.~lim.}&& \multicolumn{1}{c}{[keV]}  \\
\hline
491.8(5)	&$88(25)$ meV&--	&~~& 491(5)	&	$61^{+11}_{-8}$ meV \\
432(2)	&$50(13)$ meV&--	&~~& 435(1)	&	$50^{+10}_{-7}$ meV\\
399(2)	&~~&$<8.6$ meV&&--	&	$<3.99~\mu$eV	\\
342(4)	&~~&$<63$ meV&&--	&	$<14.2~\mu$eV	 \\
 \multicolumn{1}{c}{--~~~~}	&~~&--	&&(311(2))&	$28^{+10}_{-8}~\mu$eV	\\
301(4)	&~~&$<11$ meV&&--	&	~~~~	 \\
 \multicolumn{1}{c}{--~~~~}	&~~&--	&&$(293(2))$&	$82^{+18}_{-21}~\mu$eV\\
281(4)	&~~&$<4.2$ meV&&--	&	$<10.1~\mu$eV	 \\
243.6(15)$^{\dag}$&~~&$<8.1$ $\mu$eV&&--	&	$<21.8~\mu$eV	\\
214(4)	&~~&$<59$ $\mu$eV&&--	&	$<10.5~\mu$eV	\\
183(4)	&~~&$<3.9$ $\mu$eV&&--	&	$<1.42~\mu$eV	\\
\hline
 \multicolumn{4}{l}{\footnotesize $^*$ See Table \ref{tab:resonances}}\\
 \multicolumn{4}{l}{\footnotesize $^{\dag}$ Endt \cite{Endt1990} }\\
  \multicolumn{4}{l}{\footnotesize $^{\ddag}$ Parikh \cite{Parikh2009} }
\end{tabular}
\end{center}
\end{table}

 \section{Results and Discussion}\label{sect:disc}

The resonance strengths of 10 states within the range of $E_r = 180-492$~keV have been measured in this work and are presented in Table~\ref{tab:wg}.  The strengths of two of these resonances, $E_r = 491.8$ and 432 keV, had been measured previously by Waanders~{\it et~al.}~\cite{Waanders1983} and are in good agreement with this work.  The resonance energies measured for these two states also agree well with the literature values \cite{Endt1990,Parikh2009}.  Experimental upper limits on $\omega\gamma$ have been determined for all of the other known states within the $0.2 - 0.4$~GK Gamow energy-windows below the previously measured $E_r=432$ keV state.  All except one ($E_r=243.6$ keV) are lower than the upper limits calculated in \cite{Parikh2009} which assumed maximal spectroscopic factors. Additionally, based on the resonance energy determinations discussed in Section~\ref{sec:Er}, two new states at $E_r = 311(2)$ and 293(2)~keV are proposed.

 \subsection{Impact on nova nucleosynthesis}
To examine the impact of these measurements on the abundance of $^{33}$S in ONe novae, four different $^{33}$S($p,\gamma$)$^{34}$Cl thermonuclear rates have been calculated. All four rates were determined using states within $E_{r} = 0-840$ keV using the narrow resonance formalism \cite{IliadisText}. Rate A is a lower limit; that is, it is calculated from the lower limits for all measured resonance strengths (see Table \ref{tab:wg} and Ref.~\cite{Endt1990}).  Neither theoretical estimates for unmeasured resonances \cite{Parikh2009} nor upper limits from the present work were included in this rate.  Rate B is a very conservative upper limit; here,  the upper limits for all measured resonance strengths as well as the estimates from \cite{Parikh2009} (which had assumed maximal spectroscopic factors) for all unmeasured resonances have been used.  Rate C again uses upper limits for all measured strengths but only includes the theoretical estimates for the five unmeasured states below $E_{r}$ = 180 keV.   The four unmeasured resonances at 463, 725, 774 and 837 keV are not included in rate C.  Rate D could be considered a more likely upper limit for the rate.  It was calculated as rate C, but adopts the strengths of neighbouring resonances for the four states at $E_{r}$ = 463, 725, 774, 837 keV \cite{Waanders1983}.  The implicit assumption here is that since the direct measurement of Waanders~{\it et~al.}~\cite{Waanders1983} did not report strengths for these four states, it is unlikely that any of these strengths are larger than strengths of nearby states they did in fact measure.    The direct capture component \cite{Parikh2009} is only relevant for the lower limit (rate A), where it dominates below $T = 0.06$ GK.  The similarity between rates C and D immediately suggests that measurements of the four resonances at 463, 725, 774, and 837 keV would not significantly affect the rate.  Rather, we find the unknown strengths of the low energy resonances ($E_{r}$ $< 290$ keV) should be measured to address the difference between rate A (in which they are not included) and rates C and D (where they are included with with upper limits from the present work or maximum theoretical contributions).  Of course, a verification of the above assumption regarding the sensitivity of the Waanders~{\it et~al.}\ measurement would be welcome.  

\begin{figure}
\begin{center}  
\includegraphics[width=9 cm]{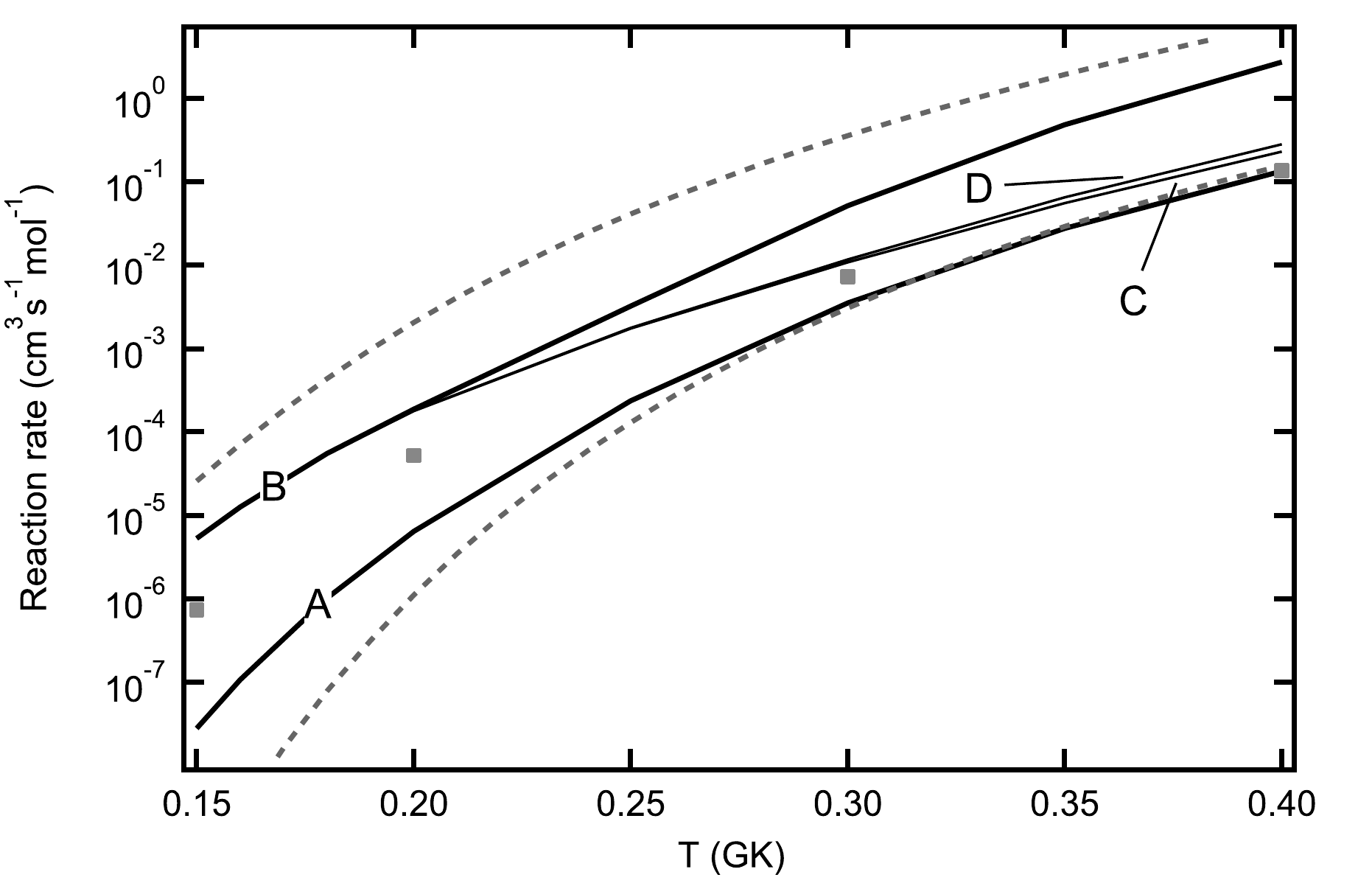}
  \caption{Thermonuclear $^{33}$S(p,$\gamma$)$^{34}$Cl rates over typical nova peak temperatures. Solid lines indicate rates calculated from this work (see text).  The dotted lines indicate the upper and lower rates from \cite{Parikh2009}. A rate calculated using a statistical model \cite{Iliadis2001,Rauscher2000} is also included for comparison (filled grey squares).   Note that this statistical model rate has been normalized to the experimental rate at T = 2 GK (see text).}
  \label{rates_abcd}
\end{center}
\end{figure}

These rates are shown in Fig.~\ref{rates_abcd} over typical peak temperatures in classical novae.  The upper and lower limits from  \cite{Parikh2009} and rates determined from a Hauser-Feshbach statistical model \cite{Iliadis2001,Rauscher2000}  are also included for comparison. This theoretical rate has been normalized to the experimental rate at $T = 2$~GK, at which the Gamow window extends over the range $E_r = 690 - 1750$~keV.  Note that Ref.~\cite{Endt1990} lists strengths for 55 known $^{33}$S($p,\gamma$)$^{34}$Cl resonances over $E_r = 430 - 1940$~keV.  As such, the experimental rate at 2~GK is essentially independent of the uncertainties on the low-energy resonances of interest in novae. At the temperatures of concern, the normalized statistical model rate is in good agreement with the bounds presented by our rates A and D.

The effect of the experimental rates from this work on nova yields was tested through a series of 1-D, hydrodynamic simulations performed with the SHIVA code \cite{Jose1998}.  Four models of 1.35~M$_{\odot}$ ONe white dwarfs, accreting H-rich material from the stellar companion at a rate of $2 \times 10^{-10}$~M$_{\odot}$ yr$^{-1}$ have been computed, with identical input physics except for the prescription adopted for the $^{33}$S($p,\gamma$)$^{34}$Cl rate.  Due to the lack of sufficient information on the relative population of the metastable and ground states in $^{34}$Cl, we have only included $^{34}$Cl as a single species in the reaction network.   All four models result in the same peak temperature (0.31 GK) and amount of mass ejected ($4.6 \times 10^{-6}$ M$_{\odot}$).  For elements with $A<33$, the nucleosynthesis accompanying the explosion did not vary between the four models, as expected.  Differences appear in the S -- Ca mass region, as shown in Table \ref{nova_yields}.  As expected, the model using rate A (our lower limit) produced the largest amount of $^{33}$S and the least amount of material at higher masses; the models using larger rates ejected less $^{33}$S and more material at higher masses.  Note that yields from the models using rates C and D did not differ within the two digit precision used in Table \ref{nova_yields}, as expected given the similarity of these two rates (see Fig. \ref{rates_abcd}).  

\begin{table}[tb]
  \caption{Mean composition of the ejecta (S -- Ca, as mass fractions) from 1.35 M$_{\odot}$ ONe white dwarf nova models adopting different $^{33}$S($p,\gamma$)$^{34}$Cl rates (see text).}  \label{nova_yields}
  \begin{center}
    \begin{tabular}{cccc}
    \hline
    Nuclide&Rate A&Rate B&~Rate C or D~\\
   \hline
$^{32}$S&~ $5.3\times10^{-2}$ ~&~$5.3\times10^{-2}$~ &$5.3\times10^{-2}$\\
$^{33}$S&~ $5.0\times10^{-4}$~ &~$6.2\times10^{-5}$ ~&$1.2\times10^{-4}$\\
$^{34}$S&~ $4.5\times10^{-4}$ ~&~$5.9\times10^{-4}$ ~&$6.0\times10^{-4}$\\
$^{35}$Cl&~ $5.4\times10^{-4}$ ~&~$7.4\times10^{-4}$~ &$7.0\times10^{-4}$\\
$^{37}$Cl&~ $2.0\times10^{-4}$ ~&~$2.7\times10^{-4}$~ &$2.5\times10^{-4}$\\
$^{36}$Ar&~ $7.4\times10^{-5}$~ &~$1.0\times10^{-4}$~ &$9.6\times10^{-5}$\\
$^{38}$Ar&~ $2.6\times10^{-5}$~ &~$3.1\times10^{-5}$~ &$3.0\times10^{-5}$\\
$^{38}$K&~ $2.8\times10^{-6}$ ~&~$4.2\times10^{-6}$~ &$3.6\times10^{-6}$\\
$^{39}$K&~ $6.1\times10^{-6}$~ &~$6.4\times10^{-6}$~ &$6.3\times10^{-6}$\\
$^{40}$Ca&~ $3.1\times10^{-5}$ ~&~$3.1\times10^{-5}$ ~&$3.1\times10^{-5}$\\
   \hline
  \end{tabular}
\label{summary}  
  \end{center}
\end{table}

The overproduction factor of $^{33}$S relative to solar varies between $\approx 20 - 150$ in our four models (using $X$($^{33}$S$_{\odot}$)~=~3.2$\times$10$^{-6}$~\cite{Piersanti2007}).  The $^{32}$S/$^{33}$S isotopic ratio, which is perhaps more relevant from the perspective of identifying nova signatures in grains, varies between $\approx 100 - 850$ when results from all four of our models are considered (the solar value is $\approx 120$~\cite{Piersanti2007}).  Considering instead only our lower rate limit (rate A) and the more likely upper limit (rate D), we obtain a range of $\approx 100 - 450$ for the $^{32}$S/$^{33}$S ratio.  This is a significant reduction from the previous estimate of $\approx 1160$ \cite{Parikh2009}.

If future measurements of sulphur isotopic ratios require better constraints for predicted ratios, our results indicate that the priority should be to reduce the upper limits of the strengths of the resonances with $180 < E_r < 290$~keV.  The contributions from these resonances dominate the difference between rates A and D in Fig.~\ref{rates_abcd}. Secondary goals would be to measure the strengths of $^{33}$S($p,\gamma$)$^{34}$Cl resonances below E$_{r}$ = 180 keV and verify that the strengths of the resonances at 463, 725, 774, and 837 keV are below $\approx 80$~meV.  While we have not examined the production of the ground versus the metastable state production of $^{34m}$Cl, the maximum variation in $^{33}$S production from our four models is only a factor of $\approx 8$.  Even with the highest $^{33}$S($p,\gamma$)$^{34}$Cl rate  leading to the highest potential $^{34m}$Cl production, it is unlikely that prospects for detecting $\gamma$-rays following the decay of any $^{34m}$Cl produced in ONe nova explosions will have improved~\cite{Parikh2009,Jose2001,Coc2000}.

\section{Conclusions}

Using the DRAGON recoil separator we have measured or determined upper limits for 7 previously unmeasured resonance strengths in $^{33}$S($p,\gamma$)$^{34}$Cl and confirmed two previously measured values at $E_r= 431$ and 492~keV.  Additionally, two new states within the $0.2-0.4$~GK Gamow window are proposed along with measurements of their resonance strengths.  From this work new upper and lower limits on the rate of the $^{33}$S($p,\gamma$)$^{34}$Cl reaction as a function of temperature were determined.  These rates were used in a series of 1-D hydrodynamic nova simulations to determine the resulting $^{33}$S abundances in a 1.35~M$_{\odot}$ ONe nova.  This results in the $^{32}$S/$^{33}$S ratio, of interest in presolar grain classification, being reduced to a likely range of $100-450$ as compared to the previous estimate of $\approx 1160$~\cite{Jose2001,Parikh2009}.  However, this range of $^{32}$S/$^{33}$S ratio still does not exclude the solar  $^{32}$S/$^{33}$S ratio of $\approx 120~$\cite{Piersanti2007}.  Further measurements, particularly a reduction of the upper limits on the resonance strengths between $180 < E_r < 290 $~keV are, therefore, recommended.

\begin{acknowledgments}
We would like to thank the beam delivery and ISAC operations groups at TRIUMF and also gratefully acknowledge the invaluable assistance in beam production from K. Jayamanna.

The authors gratefully acknowledge funding from the Natural Sciences and Engineering Research Council of Canada.  AP and JJ were partially supported by the Spanish MICINN grants AYA2010-15685 and EUI2009-04167, the Government of Catalonia grant 2009SGR-1002, the E.U. FEDER funds, and the ESF EUROCORES Program EuroGENESIS.  AP was also supported by the DFG cluster of excellence "Origin and Structure of the Universe". Authors from the USA wish to thank both the Department of Energy and the National Science Foundation for their support.  Authors from the UK would like to aknowledge the support of the Science and Technology Funding Council.  Authors from the Collaborators from CIAE wish to thank the National Natural Science Foundation of China (Grant No. 11021504) and the 973 program of China (Grant No. 2013CB834406).


\end{acknowledgments}

\bibliography{33S}

\end{document}